\documentclass[10pt, conference, letterpaper]{IEEEtran}
\IEEEoverridecommandlockouts

\usepackage{cite}
\usepackage{amsmath,amssymb,amsfonts}
\usepackage{bm} 
\usepackage{algorithm}
\usepackage{algorithmic}
\usepackage{graphicx}
\usepackage{textcomp}
\usepackage{xcolor}
\usepackage{siunitx}
\usepackage{hyperref}
\usepackage{tablefootnote}
\usepackage{multirow}
\usepackage{enumitem}
\usepackage{tikz}

\usetikzlibrary{patterns}

\newcommand{\N}{\mathcal{N}}

\newcommand{\Ecal}{\mathcal{E}}

\newcommand{\alg}[1]{\textbf{\texttt{#1}}}
\newcommand{\viola}[1]{\textcolor{red}{\textbf{#1}}}
\newcommand{\satisfy}[1]{\textcolor{blue}{\textbf{#1}}}

\begin{document}
\title{Asynchronous Risk-Aware Multi-Agent Packet Routing for Ultra-Dense LEO Satellite Networks
}
\author{}
\author{
    \IEEEauthorblockN{
        Ke He\IEEEauthorrefmark{2}\IEEEauthorrefmark{4},,
        Thang X. Vu\IEEEauthorrefmark{2}\IEEEauthorrefmark{4},,
        Le He\IEEEauthorrefmark{3},
        Lisheng Fan\IEEEauthorrefmark{3},
        Symeon Chatzinotas\IEEEauthorrefmark{2} and
        Bj\"orn Ottersten\IEEEauthorrefmark{2}
    }
    \IEEEauthorblockA{
        \IEEEauthorrefmark{2}The Interdisciplinary Centre for Security, Reliability and Trust (SnT), University of Luxembourg, L-1855 Luxembourg.
    }
    \IEEEauthorblockA{
        \IEEEauthorrefmark{3}School of Computer Sciense and Cyber Engineering, Guangzhou University, Guangzhou, China. 
    }
    \IEEEauthorrefmark{4}Corresponding authors: Ke He and Thang X. Vu (\{ke.he, thang.vu\}@uni.lu)
}

\maketitle

\begin{abstract}
The rise of ultra-dense LEO constellations creates a complex and asynchronous network environment, driven by their massive scale, dynamic topologies, and significant delays. This unique complexity demands an adaptive packet routing algorithm that is asynchronous, risk-aware, and capable of balancing diverse and often conflicting QoS objectives in a decentralized manner. However, existing methods fail to address this need, as they typically rely on impractical synchronous decision-making and/or risk-oblivious approaches. To tackle this gap, we introduce \alg{PRIMAL}, an event-driven multi-agent routing framework designed specifically to allow each satellite to act independently on its own event-driven timeline, while managing the risk of worst-case performance degradation via a principled primal-dual approach. This is achieved by enabling agents to learn the full cost distribution of the targeted QoS objectives and constrain tail-end risks. Extensive simulations on a LEO constellation with $1584$ satellites validate its superiority in effectively optimizing latency and balancing load. Compared to a recent risk-oblivious baseline, it reduces queuing delay by over $70$\%, and achieves a nearly $12$ ms end-to-end delay reduction in loaded scenarios. This is accomplished by resolving the core conflict between naive shortest-path finding and congestion avoidance, highlighting such autonomous risk-awareness as a key to robust routing.
\end{abstract}


\section{Introduction}
The rapid development of Low Earth Orbit (LEO) satellite mega-constellations is driving a new era of global connectivity \cite{comsur/KodheliLMSSMDSC21, comsur/AlHraishawiCKLC23, cm/LiWWL25}. These networks construct a dynamic space-based backbone of thousands of satellites interconnected by Inter-Satellite Links (ISLs) at altitudes of 500 to 2000 kilometers. Commercial constellations from providers like Starlink, OneWeb, Kuiper, Qianfan and Telesat are now being actively deployed with the ultimate goal of providing ubiquitous high-bandwidth and low-latency internet access to every corner of the globe\cite{jsac/ZhangDLNKXJK24}. In order to realize this vision, a core challenge  is designing a robust and adaptive packet routing mechanism for this dynamic infrastructure \cite{cm/LiWWL25}. 

However, designing such an algorithm is challenging due to the massive scale, dynamic topology, and significant propagation delays inherent in LEO networks \cite{cm/LiWWL25, cm/YangHLHL25, He2025spatiotemporal}. These characteristics make centralized control with a timely global network view impractical \cite{comsur/KodheliLMSSMDSC21, tvt/HanXZWCY23}. Consequently, an effective routing algorithm must be decentralized, allowing each satellite to operate \textit{asynchronously} and \textit{independently} using only local information. Furthermore, imbalanced global traffic distribution causes unpredictable congestion, demanding a mechanism for \textit{asynchronous risk-aware packet routing} \cite{comsur/AlHraishawiCKLC23, cm/LiWWL25, cm/YangHLHL25}. Such a mechanism should be able to manage conflicting Quality of Service (QoS) objectives, like latency minimization and load balancing, in a decentralized manner. Addressing this need is the main focus of our work.

\textbf{Related Works \& Limitations.}~ Prior work has explored both traditional and learning-based strategies \cite{lv2019research, comsur/AlHraishawiCKLC23}. Traditional rule-based methods often use static topology snapshots or geographic principles to pre-calculate routes \cite{hpsr/ZhengWTWY21, ekici2002distributed, ton/LiWW24, mobicom/LiLLLCZWWLL24}. While some approaches avoid full global knowledge \cite{jsac/ChenYZWZC24}, they are fundamentally \textit{risk-oblivious} and fail to handle dynamic events like traffic congestion \cite{tnse/HuangFTDYZ24}. To address these limitations, data-driven Deep Reinforcement Learning (DRL) has emerged as a promising direction \cite{tcom/WeiFGLLMG25, vtc/ZuoWYHJ21, jsac/LyuHFLAM24, network/ZhangDLNKSSP24, wcnc/LiWW25, lozano2025continual}. While early single-agent RL frameworks exist \cite{tcom/WeiFGLLMG25}, they suffer from scalability bottlenecks, shifting the focus to decentralized Multi-Agent Reinforcement Learning (MARL) where each satellite acts based on its local observation \cite{vtc/ZuoWYHJ21, jsac/ZhangXLFNL23, jsac/LyuHFLAM24, wcnc/LiWW25}.

However, many existing MARL methods are often misaligned with the physical reality of LEO satellite networks. Approaches based on cooperative MARL paradigms like MAPPO \cite{wcnc/LiWW25, jsac/LyuHFLAM24} typically enforce action-synchronization by discretizing time into synchronous time-slots, where agents are constrained to make at most one decision per time slot. While individual satellites can achieve high-precision physical clock synchronization, this does not resolve the impracticality of this time-slotted decision paradigm. The core issue is that forcing the naturally event-driven packet routing decision process (i.e., asynchronous packet arrivals or departures) into a rigid time-stepped joint-action model introduces artificial delays and severe scalability bottlenecks, as all agents must wait for a ``global tick'' before acting \cite{tvt/HanXZWCY23, liang2022asm}. Motivated by this point, asynchronous MARL approaches such as continual DRL with Federated Learning (FedL) and asynchrnous QMIX were proposed to address the issue \cite{lozano2025continual, cuongle2024}. However, the absence of synchronized cooperation of all agents can lead to conflicting decisions among agents. This highlights the need for a framework that can manage the risks arising from uncoordinated decentralized actions made by independent agents.

Existing attempts at risk-awareness often rely on heuristic reward shaping. This approach incorporates risk-awareness by engineering complex
weighted reward functions that try to balance objectives such as energy budgeting, latency minimization and load balancing \cite{tcom/WeiFGLLMG25, twc/HeVHNCO24, wcnc/LiWW25, infocom/SongJLPWJM23}. However, such methods lack formal guarantees, and more critically, they require extensive human-effort on trial-and-error based adjusting \cite{paternain2019constrained}. A more principled approach is Constrained Reinforcement Learning (CRL). However, recent CRL-based approaches for satellite packet routing like \cite{jsac/LyuHFLAM24}, while using a primal-dual method, were risk-myopic to constraining only the average values (neglecting tail-end risks) and relied on centralized coordinators~\cite{yang2021wcsac}, reintroducing the aforementioned scalability and synchronization problems.

\textbf{Motivations \& Contributions.}~ Our work is motivated by the two critical gaps in existing research. First, the dominant synchronous paradigm in many MARL routing algorithms is poorly suited to LEO networks. They rely on cooperative MARL frameworks that require a centralized coordinator and action-synchronization across all LEOs, which contradicts the inherently asynchronous and event-driven nature of packet routing \cite{nips/XiaoTA22, liang2022asm, tits/MendaCGBTKW19, atal/YuYGCLLXHYWW23}. Independent MARL approaches, such as \cite{lozano2025continual}, remove the need for coordination, but they risk performance degradation as agents may repeatedly interfere with each other's policies. This makes it more difficult to manage diverse and conflicting QoS objectives, reinforcing the need for an asynchronous risk-aware MARL framework. 

Second, existing strategies lack robust risk awareness. Many are \textit{risk-oblivious} \cite{tcom/WeiFGLLMG25, wcnc/LiWW25, infocom/SongJLPWJM23}, relying on heuristic reward shaping that lacks formal guarantees and significant human-efforts on coefficient adjusting~\cite{paternain2019constrained}. Others are \textit{risk-myopic} \cite{jsac/LyuHFLAM24}, optimizing only for average performance while ignoring high-impact tail-end events like sudden latency spikes \cite{tnn/ZhangLMLWLLY25}.

To address these limitations, we propose \alg{PRIMAL}, a novel \textit{asynchronous} and \textit{risk-aware} MARL framework. \alg{PRIMAL} uses an event-driven design that allows each satellite to act asynchronously based on their local information. Our method includes a principled primal-dual approach which learns the full distribution of the interested risk metrics, and directly constrains worst-case performance risks via distributional reinforcement learning. To summarize this work, our core contributions are:
\begin{itemize}[leftmargin=*, label=\scalebox{1.25}{$\bullet$}]
    \item We formulate the packet routing problem in satellite networks as an asynchronous MARL problem based on an \textbf{event-driven semi-Markov decision process}. Our model operates in continuous time, where each satellite agent acts independently and asynchronously based on its local event-driven timeline and observations. This approach eliminates the unrealistic synchronization and discrete time-step assumptions of prior RL-based routing methods, leading to a more realistic and efficient routing paradigm.
    \item We propose a \textbf{principled risk-aware routing algorithm using a distributional primal-dual framework}. Instead of learning only the expected costs, our agents learn the full conditional distribution of routing outcomes via quantile regression. By directly constraining the Conditional Value-at-Risk (CVaR) at a given risk level, our algorithm can effectively mitigate tail-end risks, ensuring the routing policy is robust against worst-case performance degradation.
    \item We develop a \textbf{decentralized and synchronization-free scalable learning architecture} without reliance on a centralized coordinator that needs action-synchronization and global state information. This enables extensions such as online federated learning \cite{lozano2025continual} that facilitates scalable and practical online adaption in real-world mega-constellations.
\end{itemize}

\section{System Model \& Problem Formulation}\label{sec:system_model}
\subsection{Network Model}
\begin{figure}[t!]
    \centering
    \includegraphics[width=0.7\linewidth]{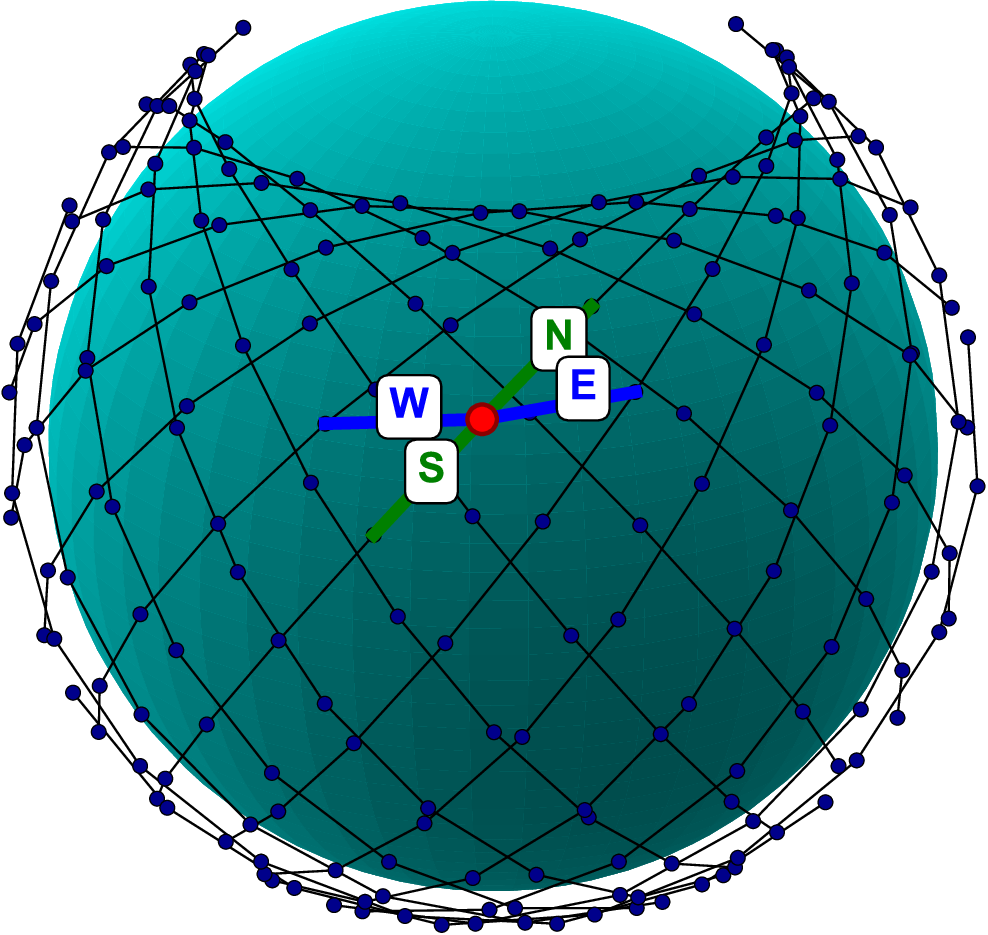}
    \caption{Illustration of a Walker-Delta LEO constellation with grid topology, where each satellite connects two intra-plane neighbors (N+S) and two inter-plane neighbors (W+E). Note that we define the four directions w.r.t. the rotating direction of the orbit. }
    \label{fig:system_model}
\end{figure}

As illustrated in Fig. \ref{fig:system_model}, we model the LEO satellite network as a time-varying directed graph $\mathcal{G}_t = (\N, \Ecal(t))$, where the node set $\N$ consists of satellites $\mathcal{N}_s$ and ground stations $\mathcal{N}_g$.

\subsubsection{Nodes}
The $|\mathcal{N}_s|$ satellites are arranged in a Walker constellation, forming a grid-like topology with Inter-Satellite Links (ISLs) \cite{cm/LiWWL25}. Each satellite acts as a store-and-forward router with finite buffers. The $|\mathcal{N}_g|$ ground stations act as traffic entry/exit points, connecting to the network via Ground-to-Satellite Links (GSLs) with their respective \textit{access satellites}.

\subsubsection{Links, Packets, and Queues}
The edge set $\Ecal(t)$ includes dynamic ISLs with Free Space Optical (FSO) Lasers, and GSLs with Ka-Band radio frequency (RF) links, whose availability depends on line-of-sight (LoS) and minimum elevation angles \cite{cakaj2021parameters}. Data is transmitted as packets, each defined by a tuple $p \triangleq \{s_p, d_p, L_p, \tau_{p}, \tau^{ttl}_p \} \in \mathcal{P}$, representing its source, destination, size, creation time, and a Time-to-Live (TTL) field initialized to the maximum TTL $H$, and $\mathcal{P}$ denotes the set of packets. Packets are processed in a First-In-First-Out (FIFO) manner and are dropped if TTL expires or if buffers are full.

\subsection{Communication and Delay Model}
The end-to-end (E2E) delay for a packet traversing the network is the sum of propagation, transmission, and queuing delays at each hop. To model these delays, we begin by defining a single, hop-indexed binary decision variable $x_{p, ij}^h=1$ if packet $p \in \mathcal{P}$ traverses the link $(i, j)$ as its $h$-th hop in its path and $0$ otherwise, where $h \in \{0, 1, ..., H-1\}$. The delay at hop $h$ depends on the packet's arrival time at the starting node of the hop, and the arrival time is determined by the sum of delays from all previous hops ($0$ to $h-1$). For a packet $p$ traversing a link $(i, j)$ at time $t$, the single-hop delay is
\begin{align}
    D^h_{p, ij} = D^{P}_{ij}(\tau^h_{p}) + D^{T}_{ij}(L_p, \tau^h_{p}) + D^{Q}_{ij}(\tau^h_{p}),
\end{align}
where $D^{P}_{ij}(t)$ is the propagation delay, $D^{T}_{ij}(L_p, t)$ is the transmission delay, and $D^{Q}_{ij}(t)$ is the queuing delay of the sending node $i$ at arrival time $t = \tau^h_{p}$. The arrival time $\tau^h_{p}$ at the start of hop $h$ can be recursively defined as
\begin{align}
    \tau^h_{p} = \tau_{p} + \sum_{k=0}^{h-1} \sum_{(u,v) \in \Ecal(\tau_{p,k})} x_{uv,k}^p \cdot D^p_{uv,k}
\end{align}
Now we are ready to introduce the models of each delay component.

\subsubsection{Propagation Delay}
The propagation delay $D^{P}_{ij}(t)$ is the time needed for a signal to travel from node $i$ to node $j$ via link $(i, j)$ at time $t$. It is determined by the Euclidean distance $d_{ij}(t)$ between the nodes at time $t$, as well as the speed of light $\kappa_c$, which is given by
\begin{equation}
    D^{P}_{ij}(t) = \frac{d_{ij}(t)}{\kappa_c}.
\end{equation}

\subsubsection{Transmission Delay}
The transmission delay $D^{T}_{ij}(L_p, t)$ is the time required to send all the $L_p$ bits of packet $p$ onto communication link $(i,j)$ at time $t$. It is a function of the packet size $L_p$ and the link's achievable data rate $R_{ij}(t)$. For GSLs operating in the Ka-Band, the data rate is modeled by the Shannon-Hartley theorem, which depends on the link's bandwidth $B_{ij}$ and its Signal-to-Noise Ratio (SNR):
\begin{equation}
    R_{ij}^{GSL}(t) = B_{ij} \log_2(1 + \text{SNR}_{ij}(t)).
\end{equation}
The $\text{SNR}_{ij}(t)$ is determined by \cite{wcnc/LiWW25}:
\begin{equation}
    \text{SNR}_{ij}(t) = \frac{P^T_{ij} G^T_{ij} G^R_{ij}}{L_{ij}(t) \kappa_B T_{ij} B_{ij}},
\end{equation}
where $P^T_{ij}$ is the antenna transmission power, $G^T_{ij}$ and $G^R_{ij}$ are the transmitter and receiver antenna gains, $\kappa_B$ is the Boltzmann’s constant, $L_{ij}$ is the Free Space Path Loss (FSPL) and $T_{ij}$ is the system noise temperature in Kelvin. $L_{ij}$ is a function of the distance $d_{ij}(t)$ and carrier frequency of Ka-Band $f_c$:
\begin{equation}
    L_{ij}(t) = \left(\frac{4\pi d_{ij}(t) f_c}{\kappa_c}\right)^2.
\end{equation}

For FSO Laser based ISLs, which rely on FSO communication, the data rate is modeled differently \cite{jsac/ChenYZWZC24}:
\begin{equation}
    R_{ij}^{ISL}(t) = \frac{\tilde{B}_{ij}}{2} \log_2\left(1 + \kappa_1 \cdot e^{-\kappa_2 \cdot d_{ij}(t)}\right),
\end{equation}
where $\tilde{B}_{ij}$ is the optical bandwidth. The parameters $\kappa_1$ and $\kappa_2$ are related to the average optical SNR and attenuation conditions \cite{tvt/LeePBK23, comsur/ChaabanRA23}. Consequently, the transmission delay for a packet $p$ of size $L_p$ over link $(i,j)$ is given by
\begin{equation}
    D^{T}_{ij}(L_p, t) = \frac{L_p}{R_{ij}(t)},
\end{equation}
where $R_{ij}(t)$ is either $R_{ij}^{GSL}(t)$ or $R_{ij}^{ISL}(t)$ depending on the link type.

\subsubsection{Queuing Delay}
The queuing delay $D^{Q}_{ij}(t)$ is the time a packet spends waiting in an output buffer before its transmission begins. In a FIFO output queue, this delay is the sum of the transmission delays of all preceding packets in the queue for the same outgoing link. If packet $p$ arrives at node $i$ at time $t$ and is routed to next-hop $j$, its queuing delay can be approximately calculated as
\begin{equation}
    D^{Q}_{ij}(t) = \sum_{q \in \mathcal{P}_{ij}(t)} \frac{L_q}{R_{ij}(t)},
\end{equation}
where $\mathcal{P}_{ij}(t)$ is the set of packets already in the output queue for link $(i, j)$ at the time $t$. The queuing delay is a direct indicator of local congestion, as a congested node inherently results in longer packet queue. Consequently, the accumulated queuing delay of a packet's journey can serve as an ideal metric for evaluating the network's load balancing performance.

\subsection{Problem Formulation}
The packet routing problem can be formulated as a large-scale non-linear integer programming problem. The main objective is to find an optimal routing policy, defined by the set of $\vert \mathcal{P} \vert \times \vert \mathcal{E} \vert \times H $ decision variables $\{x_{p, ij}^h\}$, that minimizes the total E2E delay of each packet
\begin{align}
   D_p = \sum_{h=0}^{H-1} \sum_{(i,j) \in \Ecal({\tau^h_{p})}} x_{p, ij}^h \cdot D^h_{p, ij},
\end{align}
while subjects to several fundamental constraints that define valid routing paths.

First, a valid path must be contiguous. The destination node of any given hop must serve as the source node for the subsequent hop. This path connectivity constraint is enforced for all non-terminal nodes:
\begin{align}
\sum_{i \in \N_k} x_{ik,p}^h = \sum_{j \in \N_k} x_{kj,p}^{h+1} \quad \forall p \in \mathcal{P}, k \in \mathcal{S}, h < H-1. \tag{C1}
\end{align}
In addition, each packet's journey must start at its source ground station $s_p$ and eventually terminate at its destination ground station $d_p$. The source and destination constraints for packet delivery are
\begin{align}
\sum_{j \in \N_{s_p}} x_{s_p,j,p}^1 = 1 \quad \text{and} \quad \sum_{h=0}^{H-1} \sum_{i \in \N_{d_p}} x_{i,d_p,p}^h = 1 \quad \forall p \in \mathcal{P}. \tag{C2}
\end{align}
Moreover, to ensure that the path is simple and loop-free within a hop index, a packet can traverse at most one link for any given hop $h$:
\begin{align}
\sum_{(i,j) \in \Ecal({\tau^h_{p}})} x_{p, ij}^h \le 1 \quad \forall p \in \mathcal{P}, h \in \{0,1, ..., H-1\} \tag{C3}
\end{align}
To manage congestion, we add a constraint on the maximum accumulated queuing delay:
\begin{align}
\label{eq:c4}
D^Q_p = \sum_{h, (i,j)} x_{p, ij}^h \cdot D^{Q}_{ij}(\tau^h_{p}) \leq D^Q_{max}, \quad \forall p \in \mathcal{P} \tag{C4}
\end{align}
where $D^Q_{max}$ is a predefined threshold. The full problem is:
\begin{align}
    \textbf{P1:} \min_{\{x_{p, ij}^h\}} \quad \frac{1}{\vert \mathcal{P} \vert}\sum_{p \in \mathcal{P}} D_p  \quad \text{s.t.} ~~ \text{(C1), (C2), (C3), (C4)} 
    \end{align}
This problem is intractable for decentralized solutions due to its non-linear and interdependent nature (e.g., queuing delay couples all routing decisions) and the massive scale of the network, which makes centralized solvers infeasible \cite{fonoberova2005optimal}. This motivates our decentralized and scalable learning framework.
\section{Principled Risk-Aware Independent Multi-Agent Learning}
\label{sec:risk_aware_marl}
To overcome the limitations of synchronized MARL models discussed earlier \cite{jsac/LyuHFLAM24, wcnc/LiWW25}, we adopt an asynchronous event-driven perspective. We model the trajectory of a single packet as a \textit{Partially-Observed Constrained Semi-Markov Decision Process} (POCSMDP), where satellite routers are independent decision-makers. This packet-centric view defines a finite-horizon learning episode for each packet's journey.

\subsection{Event-Driven Semi-Markov Decision Process}
The POCSMDP for a packet $p$ is defined by the tuple $\langle \mathcal{S}, \mathcal{A}, \mathcal{T},  r, \{c_k\}, \mathcal{O}, H, \gamma_r, \gamma_c \rangle_p$, where:
\begin{itemize}
    \item  $\mathcal{S}$ is the global network state space. A state $s_h \in \mathcal{S}$ is a snapshot of the network's physical status and packet-specific information at the $h$-th hop.
    \item $\mathcal{A}$ is the packet routing action space at a satellite, i.e., the set of four outgoing ISLs (NSWE).
    \item $\mathcal{T}(s^{\prime}, \tau, | s, a)$ is the transition probability to state $s^{\prime}$ after a variable duration $\tau$ (the single-hop delay), making this a \textit{semi-Markov} process.
    \item $\mathcal{O}$ is the observation function yielding a local observation $o \sim \mathcal{O}(\cdot|s)$, which includes packet state and local node/neighbor statistics, making the process \textit{partially observable}.
    \item $r(o, a, o^\prime)$ and $\{c_k(o, a, o^\prime)\}_{k=1}^K$ are the reward (primary objective) and QoS cost functions (e.g., load balancing) for a given state transition.
    \item $\gamma_r, \gamma_c \in (0, 1]$ are discount factors.
\end{itemize}
This event-driven formulation ensures that satellites react to packet arrivals in real-time based on current local information. While each packet defines a conceptual learning episode, the physical agents are the satellites. For scalability, we employ parameter sharing, where all satellites use a single, homogeneous policy $\pi_{\theta}(a|o)$, stored locally but shared across the network, enabling distributed and asynchronous execution.

\subsection{Maximum Entropy Constrained Reinforcement Learning}
Following the event-driven POCSMDP framework, we can solve the routing problem \textbf{P1} via CRL \cite{nips/JungCPS22}. Since each satellite acts independently without coordination, maintaining a certain level of policy stochasticity is beneficial for exploration and for mitigating the non-stationary environment issue arising from the concurrent learning of other agents. This can be achieved by incorporating an additional constraint on the expected policy entropy during learning, known as maximum entropy RL \cite{haarnoja2018soft}. For each packet $p$, we are interested in the two variables, the reward-return $Z^{r}_{\pi} = \sum_{h=0}^{H-1} \gamma_r^h r(o_h, a_h, o_{h+1})$, and the cost-return $Z^{c_k}_{\pi} = \sum_{h=0}^{H-1} \gamma_c^h c_k(o_h, a_h, o_{h+1})$, all induced by a fixed policy $\pi$. The overall objective is to find the optimal policy that solves the following constrained optimization problem:
\begin{subequations}
\begin{align}
\textbf{P2:} \quad \max_{\pi} \quad & \mathcal{J}_r(\pi) = \mathbb{E} \left[ Z^r_{\pi} \right]  \\
\text{s.t.} \quad & \mathcal{J}_{c_k}(\pi) \leq D_k, \quad \forall k \in \{1, \ldots, K\} \\
\quad & \mathcal{H}\left(\pi(\cdot \vert o_h)\right) \geq \bar{\mathcal{H}}, \quad \forall h \in \{1, \ldots, H\}
\end{align}
\end{subequations}
where $\mathcal{H}\left(\pi(\cdot|o_h)\right) = \mathop{\mathbb{E}}_{a \sim \pi(\cdot|o)}\left[ -\log \pi(a \vert o)\right]$ denotes the entropy of the policy, and $\bar{\mathcal{H}}$ is the desired minimum expected policy entropy. The term $\mathcal{J}_{C_k}(\pi) \leq D_k$ represents a placeholder for the $k$-th QoS constraint, which can be a constraint on the expected cost-return as
\begin{align}\label{eq:constraint_avg}
   \mathcal{J}_{c_k}(\pi) = \mathbb{E} \left[ Z^{c_k}_{\pi} \right] \leq D_k,
\end{align}
where $D_k$ is the desired cost threshold. 

In addition to constraining the expected costs, we further consider risk-averse probabilistic constraints based on Conditional Value-at-Risk ($\mathrm{CVaR}$). Specifically, the Value-at-Risk ($\mathrm{VaR}$) at risk level $\epsilon_k \in (0, 1) $ is the $(1-\epsilon_k)$-quantile of the cost distribution defined as
\begin{align}
    \mathrm{VaR}_{\epsilon_k}(Z^{c_k}_{\pi}) = \inf\{z \in \mathbb{R} : F_{Z^{c_k}_{\pi}}(z) \ge 1-\epsilon_k\},
\end{align}
where $F_{Z^{c_k}_{\pi}}$ is the Cumulative Distribution Function (CDF) of $Z^{c_k}_{\pi}$. The $\mathrm{CVaR}$ is then defined as the expected cost in the worst $\epsilon_k$ tail of the distribution \cite{tnn/ZhangLMLWLLY25}:
\begin{align}
    \mathrm{CVaR}_{\epsilon_k}(Z^{c_k}_{\pi}) = \mathbb{E}_{\pi}\bigg[Z^{c_k}_{\pi} \big\vert Z^{c_k}_{\pi} \ge \mathrm{VaR}_{\epsilon_k}(Z^{c_k}_{\pi})\bigg],
\end{align}
Then, $\mathcal{J}_{c_k}(\pi) \leq D_k$ can be a risk-averse constraint as
\begin{align}\label{eq:constraint_cvar}
    \mathcal{J}_{c_k}(\pi) = \mathrm{CVaR}_{\epsilon_k}\left(Z^{c_k}_{\pi}\right) \leq D_k,
\end{align}
which requires that the expected cost-return $Z^{c_k}_{\pi}$, conditioned on being in the worst $\epsilon_k$-percentile of outcomes, does not exceed the threshold $D_k$. This allows for direct control over worst-case scenarios, moving beyond simple average performance.

Note that traditional methods for solving \textbf{P2} often rely on reward engineering, where a hand-crafted reward function is designed as a weighted sum of the main objective $\mathcal{J}_{r}$ and the cost components $\mathcal{J}_{c_k}$. This approach relaxes the constraints by incorporating them as penalties into the objective function, using a set of fixed coefficients. Though effective, it requires manually adjusting these coefficients to balance multiple, often conflicting objectives, which is known to be notoriously challenging in practice \cite{paternain2019constrained}.

In contrast, primal-dual learning introduces a more principled alternative that directly solves the CRL problem by learning the best multipliers for the constraints \cite{ml/YangSTS23, icml/DabneyOSM18, tnn/ZhangLMLWLLY25}.  Being reward-agnostic and requiring no prior knowledge, the approach can handle a wide range of general constraints, which eliminates the significant human-efforts required on adjusting the penalty coefficients.

Hence, we propose the \alg{P}rincipled \alg{R}isk-aware \alg{I}ndependent \alg{M}ulti-\alg{A}gent \alg{L}earning (\alg{PRIMAL}) framework to solve \textbf{P2} via primal-dual learning. Our approach first transforms the constrained problem into an unconstrained one via Lagrange multipliers. We then solve the resulting Lagrangian dual problem by extending the Soft Actor-Critic (SAC) algorithm to our multi-agent setting with discrete actions \cite{haarnoja2018soft, tmlr/ZhouWLLXS00Y24}. We now introduce two variants of our algorithm: \alg{PRIMAL-Avg}, which handles constraints on expected costs, and \alg{PRIMAL-CVaR}, which addresses constraints on worst-case costs. Both the two variants work asynchronously and independently with only limited local information, ensuring scalability and robustness in large-scale systems.

\subsection{PRIMAL-Avg: Routing with Expected Cost Constraints}
To solve the constrained optimization problem \textbf{P2} with expectation constraints, we  define the Lagrangian as:
\begin{align}
    \mathcal{L}(\pi, \bm{\lambda}, \alpha) = \mathcal{J}_r(\pi) & + \alpha \left( \mathbb{E}_{\pi}[\mathcal{H}(\pi)] - \bar{\mathcal{H}} \right)  \nonumber \\
     & -  \sum_{k=1}^K \lambda_{k} \left( \mathcal{J}_{c_k}({\pi}) - D_k \right),
\end{align}
where $\bm{\lambda} = \{\lambda_1, \ldots, \lambda_K\}$ with $\lambda_{k} \geq 0$ and $\alpha \geq 0$ are the Lagrange multipliers for the $K$ cost constraints and entropy constraint, respectively. This Lagrangian function combines the original objective with the constraints into a single equation. The core idea of the primal-dual method is to transform the original constrained problem, known as the primal problem, into an equivalent dual problem. We first define the dual function $g(\lambda, \alpha) = \max_{\pi} \min_{\bm{\lambda} \geq \bm{0}, \alpha \ge 0} \mathcal{L}(\pi, \bm{\lambda}, \alpha)$ as the maximum value of the Lagrangian with respect to the policy $\pi$ for a fixed set of multipliers. The dual problem then involves finding the multipliers that \textit{minimize} the dual function, i.e., $\min_{\lambda\ge0,\alpha\ge0} g(\lambda, \alpha)$. Note that primal-dual CRL has been proved to have strong duality for single-agent fully observable RL \cite{paternain2019constrained}. However, the constrained and partially observable MARL problem of our case is fundamentally more challenging and highly nonconvex, such that there exist none known strong duality guarantees \cite{chen2024hardness}. Despite this fact, primal-dual approach still serves as a feasible and principled way to solve the problem approximately.

We approach this using an actor-critic framework with function approximation, i.e., neural networks, for the policy (actor) and the state-action value functions (Q-functions, as the critics). In the maximum entropy framework, the soft reward critic $Q^{r}_{\phi}$, parameterized by $\phi$, is defined as the expected sum of future rewards and entropy bonuses after taking action $a$ in observation $o$ and then following policy $\pi_{\theta}$ thereafter:
\begin{align}
    Q^{r}_{\phi}(o, a)  = \mathbb{E}_{\pi_{\theta}} \bigg[ \sum_{h=0}^{H-1} & \gamma_r^{h} r(o_h, a_h, o_{h+1}) \nonumber \\
    & + \alpha \mathcal{H}(\pi(\cdot|o_h)) \big| o_0=o, a_0=a \bigg],
\end{align}
Analogously, we have the $k$-th cost critic $Q^{c}_{\psi_k}$ as
\begin{align}
    Q^{c}_{\psi_k}(o, a)  = \mathop{\mathbb{E}}_{\pi_{\theta}} \bigg[ \sum_{h=0}^{H-1} \gamma_c^{h} c_k(o_h, a_h, o_{h+1}) \big| o_0=o, a_0=a \bigg],
\end{align}
with parameters $\psi_k$. Then, we have the recursive Bellman equations as
\begin{align}
    Q^{r}_{\phi}(o, a) &= r(o,a,o^{\prime}) \nonumber\\  + & \gamma_r \mathbb{E}_{a^{\prime} \sim \pi_{\theta}(\cdot|o^{\prime})}\left[Q^{r}_{\phi}(o^{\prime}, a^{\prime}) -\alpha \log \pi(a^{\prime}|o^{\prime})\right], \label{eq:bellman_qr}\\
    Q^{c}_{\psi_k}(o, a) &= c_k(o,a,o^{\prime}) + \gamma_c \mathbb{E}_{a^{\prime} \sim \pi_{\theta}(\cdot|o^{\prime})} \left[Q^{c}_{\psi_k}(o^{\prime}, a^{\prime})\right] \label{eq:bellman_qc}
\end{align}
For discrete actions, we use $Q^r_{\phi}(o) \in \mathbb{R}^{|\mathcal{A}|\times 1}$, $Q^c_{\psi}(o) \in \mathbb{R}^{|\mathcal{A}|\times 1}$ and $\pi_{\theta}(o) \in \mathbb{R}^{|\mathcal{A}|\times 1}$ to denote the per-action outputs. The reward and cost critics can be learned by comparing their predictions to a Temporal Difference (TD) target calculated from the experience $(o, a, r, \{c_k\}, o^\prime) \sim \mathcal{D}$ drawn from a replay buffer $\mathcal{D}$. Formally, the TD targets can be computed following the Bellman equations as
\begin{align}
    y^r_{\phi^\prime} &= r + \gamma_r \pi^\top_{\theta}(o^\prime) \left(Q^{r}_{\phi^\prime}(o^{\prime}) -\alpha \log \pi(o^{\prime})\right), \label{eq:td_target_qr}\\
    y^c_{\psi^\prime_k} &= c_k + \gamma_c \pi^\top_{\theta}(o^\prime) Q^{c}_{\psi_k^\prime}(o^{\prime}) \label{eq:td_target_qc},
\end{align}
where $(\cdot)^\top$ denotes the matrix transpose. In practice, these targets are estimated via the corresponding target networks $Q^{r}_{\phi^\prime}$ and $Q^{c}_{\psi_k^\prime}$ with mirrored parameters $\phi^\prime$ and $\{\psi_k^\prime\}^K_{k=1}$. The SAC framework optimizes the reward and cost critics according to their TD errors as
\begin{align}
    \mathcal{L}_{\phi} &= \mathbb{E}_{(o, a, r, o^\prime) \sim \mathcal{D}} \left[ \frac{1}{2}\left( Q^{r}_{\phi}(o, a) - y^r_{\phi^\prime}  \right)^2 \right], \label{eq:loss_qr} \\
    \mathcal{L}_{\psi_k} &= \mathbb{E}_{(o, a, c_k, o^\prime) \sim \mathcal{D}} \left[ \frac{1}{2}\left( Q^{c}_{\psi_k}(o, a) - y^c_{\psi^\prime_k}  \right)^2 \right] \label{eq:loss_qc_avg},
\end{align}

For the policy update, the policy parameters $\theta$ are optmized by minimizing the KL-divergence between the policy and a target distribution derived from the Q-function. Specifically, the policy is updated towards the exponential of the Lagrangian-based state-action values \cite{tmlr/ZhouWLLXS00Y24}. The objective for the actor is to minimize the following loss function\cite{haarnoja2018soft}:
\begin{align}\label{eq:loss_actor_avg}
\mathcal{L}_{\theta} = \mathop{\mathbb{E}}_{o \sim \mathcal{D}} \bigg[ \pi^\top_{\theta}(o) \big(\alpha \log \pi_{\theta}(o) - Q^{r}_{\phi}(o) + \sum_k \lambda_k Q^{c}_{\psi_k}(o)\big) \bigg]
\end{align}
This update encourages the policy to select actions that have high reward Q-values and low cost Q-values (weighted by the Lagrange multipliers $\lambda_k$), while also maintaining high entropy to facilitate exploration.

The Lagrange multipliers, which are crucial for enforcing the constraints, are updated to minimize the Lagrangian. This results in simple update rules where each multiplier is adjusted based on the extent of its corresponding constraint violation.
\begin{align}
\mathcal{L}_{\lambda_k} &= \mathop{\mathbb{E}}_{\substack{o \sim \mathcal{D}}}\left[ \lambda_k \left( \pi^\top_{\theta}(o) Q^{c}_{\psi_k}(o) - D_k \right) \right], \label{eq:loss_lambda_avg}\\
\mathcal{L}_{\alpha} &= \mathop{\mathbb{E}}_{\substack{o \sim \mathcal{D}}} \left[ \alpha \left(-\pi^\top_{\theta}(o)\log \pi_{\theta}(o) - \bar{\mathcal{H}} \right) \right]. \label{eq:loss_alpha}
\end{align}
The update for $\lambda_k$ in \eqref{eq:loss_lambda_avg} increases the multiplier if the estimated cost exceeds the threshold $D_k$, thereby strengthening the penalty on costly actions in the actor's objective. Conversely, it decreases if the constraint is satisfied. Similarly, the entropy multiplier $\alpha$ is adjusted in \eqref{eq:loss_alpha} to ensure the policy's entropy remains close to the target level $\bar{\mathcal{H}}$.

\subsection{PRIMAL-CVaR: Routing with Worst-Case Cost Constraints}
While \alg{PRIMAL-Avg} ensures that QoS constraints are met on average, it remains oblivious to low-probability but high-impact events, such as sudden latency spikes that cause cascading network congestion. For mission-critical infrastructure like LEO networks, such tail-end risks are unacceptable. To address this, we introduce \alg{PRIMAL-CVaR}, a variant that directly controls the tail-end risk of the cost distribution by satisfying the $\mathrm{CVaR}$ constraints defined in \eqref{eq:constraint_cvar}. This requires moving beyond learning the mere expectation of the cost-return.

To be aware of the worst-case cost values, the critic should be able to learn the full conditional distribution of the cost-return $Z^{c_k}_{\pi}$. We achieve this using a powerful distributional RL method known as Implicit Quantile Networks (IQN) \cite{icml/DabneyOSM18}. Unlike traditional methods that learn the expected value (i.e., the mean) of a return, IQN aims to capture its full distribution. Its core idea is to approximate the quantile function (the inverse of the CDF) by learning a mapping from a probability $\zeta \in [0, 1]$ to the corresponding return value. This enables the network to implicitly model the entire distribution by being able to estimate any of its quantiles.

Specifically, we adapt IQN for discrete multi-agent SAC. This is realized by a network, $Q^{c}_{\psi_k}(o, a, \zeta)$, which takes a sample $\zeta \sim U(0,1)$ as an additional input to generate a value for that specific quantile. This technique provides a far richer and more accurate representation of the cost-return distribution, forming a solid foundation for risk-aware control \cite{ml/YangSTS23, nips/JungCPS22}. The IQN-based cost critic is trained by minimizing the quantile regression loss, guided by the distributional Bellman equation:
\begin{align}
    Z^{c_k}_{\pi}(o, a) & = \sum^{H-1}_{h=0} \gamma_c^h c_k(o_h, a_h, o_{h+1}) | o_0=o, a_0=a \nonumber \\
    &\overset{D}{=} c_k(o,a,o') + \gamma_c Z^{c_k}_{\pi}(o^{\prime}, a^{\prime}),
\end{align}
where $o' \sim P(\cdot|o, a)$, $a' \sim \pi_\theta(\cdot|o')$ and $\overset{D}{=}$ denotes equality in distribution. To leverage this for network training, we employ a specific sampling strategy for each transition $(o, a, c_k, o')$ drawn from the replay buffer $\mathcal{D}$. First, we sample $N$ quantile fractions from the standard uniform distribution, i.e., $\{\zeta_i\}_{i=1}^N \sim \mathcal{U}(0,1)$, to evaluate the current quantile estimates $Q^{c}_{\psi_k}(o, a, \zeta_i)$. Second, we sample another $N'$ quantile fractions $\{\zeta'_j\}_{j=1}^{N'} \sim \mathcal{U}(0,1)$ to construct the TD target using the target network $Q^{c}_{\psi'_k}$. For each pair $(i, j)$, the TD error is:
\begin{align}
    \delta_{ij} = \left(c_k + \gamma_c \pi^{\top}_{\theta}(o') Q^{c}_{\psi'_k}(o', \zeta'_j)\right) - Q^{c}_{\psi_k}(o, a, \zeta_i),
\end{align}
where $Q^{c}_{\psi'_k}(o', \zeta'_j) \in \mathbb{R}^{|\mathcal{A}|\times 1}$ outputs the per-action per-sample quantile Q-values for discrete action spaces. The parameters $\psi_k$ of the cost critic are then updated by minimizing the total quantile Huber loss, averaged over all samples:
\begin{align}\label{eq:loss_qc_avg_cvar}
    \mathcal{L}_{\psi_k} = \frac{1}{N'}\sum_{i=1}^N \sum_{j=1}^{N'} \mathop{\mathbb{E}} \left[ \rho_{\zeta_i}(\delta_{ij}) \right],
\end{align}
where $\rho_{\zeta_i}(u) = |\zeta_i - \mathbb{I}(u < 0)| \mathcal{L}_{\text{Huber}}(u)$ is the quantile Huber loss  \cite{icml/DabneyOSM18}. This asymmetrically weighted loss penalizes over-estimation and under-estimation differently for each quantile $\zeta_i$, which forces the critic to learn an accurate representation of the entire distribution  \cite{tnn/ZhangLMLWLLY25}.

\begin{algorithm}[t!]
\caption{The PRIMAL Algorithm}
\label{alg:primal}
\begin{algorithmic}[1]
\STATE \textbf{Input:} Risk mode $\alg{M} \in \{\alg{Avg}, \alg{CVaR}\}$, $\eta \in (0,1]$
\STATE Initialize $\pi_{\theta}$, $Q^{r}_{\phi}$, $\{Q^{c}_{\psi_k}\}_{k=1}^K$ and their targets $\phi', \{\psi_k'\}$
\STATE $\phi' \gets \phi$ and $\psi_k' \gets \psi_k$ 
\STATE Initialize multipliers $\bm{\lambda}$, $\alpha$, and shared replay buffer $\mathcal{D}$

\FOR{each event $e$}
    \IF{$e$ is a packet arrival event $(p, h)$}
        \STATE Get packet $p$ observation $o_h \sim \mathcal{O}(\cdot|s_h)$
        \STATE Execute action $a_h \sim \pi_{\theta}(\cdot|o_h)$
    \ELSIF{$e$ is an action completion event $(o',r,\{c_k\})$}
        \STATE Collect transition $(o, a, r, \{c_k\}, o')$ and add to $\mathcal{D}$
    \ENDIF
    \FOR{each training step}
        \STATE Sample a mini-batch $(o, a, r, \{c_k\}, o') \sim \mathcal{D}$.
        \STATE Update reward critic $\phi$ via \eqref{eq:loss_qr}
        \IF{$\alg{M} = \alg{Avg}$}
            \STATE Update expected cost critic $\psi_k$ via \eqref{eq:loss_qc_avg}
            \STATE Update actor $\theta$ via \eqref{eq:loss_actor_avg}.
            \STATE Update cost multipliers $\bm{\lambda}$ via \eqref{eq:loss_lambda_avg}.
        \ELSIF{$\alg{M} = \alg{CVaR}$}
            \STATE Update distributional cost critic $\psi_k$ via \eqref{eq:loss_qc_avg_cvar}
            \STATE Update actor $\theta$ via \eqref{eq:loss_actor_cvar}.
            \STATE Update cost multipliers $\bm{\lambda}$ via \eqref{eq:loss_lambda_cvar}.
        \ENDIF
        \STATE Update entropy multiplier $\alpha$ via \eqref{eq:loss_alpha}.
        \STATE Soft update reward target: $\phi' \leftarrow \eta \phi + (1-\eta) \phi'$
        \STATE Soft update cost target: $\psi_k' \leftarrow \eta \psi_k + (1-\eta) \psi_k'$
    \ENDFOR
\ENDFOR
\end{algorithmic}
\end{algorithm}

With a fully characterized cost distribution, we can directly estimate the worst-case costs. To estimate the $\mathrm{CVaR}_{\epsilon_k}$ at a given risk level $\epsilon_k$ for a state-action pair, $\Gamma_{\epsilon_k}(o,a) = \mathrm{CVaR}_{\epsilon_k}(Z^{c_k}_\pi(o,a))$, we draw $N^k$ i.i.d. samples $\{\zeta_m\}_{m=1}^{N^k}$ from the re-parametrized uniform distribution $\mathcal{U}(1-\epsilon_k, 1)$. The $\mathrm{CVaR}$ is then approximated by averaging the critic's output for these tail-end quantile fractions:
\begin{align}
    \Gamma_{\epsilon_k}(o,a) \approx \frac{1}{N^k} \sum_{m=1}^{N^k} Q^{c}_{\psi_k}(o, a, \zeta_m).
\end{align}
This estimate is then used to guide the actor, where the actor's objective function is modified to incorporate the $\mathrm{CVaR}$ estimate, averaged over the policy's action distribution:
\begin{align}\label{eq:loss_actor_cvar}
\mathcal{L}_{\theta} = \mathop{\mathbb{E}}_{o \sim \mathcal{D}} \left[ \pi^\top_{\theta}(o) \left( \alpha \log \pi_{\theta}(o) - Q^{r}_{\phi}(o) + \sum_k \lambda_k \Gamma_{\epsilon_k}(o) \right) \right],
\end{align}
where $\Gamma_{\epsilon_k}(o) \in \mathbb{R}^{|\mathcal{A}| \times 1}$ is a vector of per-action $\Gamma_{\epsilon_k}(o, a)$ estimates for discrete action space. The update for the Lagrange multiplier $\lambda_k$ is also adjusted to reflect the $\mathrm{CVaR}$ constraint violation, ensuring the policy is pushed towards risk-averse behavior:
\begin{align}\label{eq:loss_lambda_cvar}
    \mathcal{L}_{\lambda_k} = \lambda_k \mathop{\mathbb{E}}_{\substack{o \sim \mathcal{D}}} \bigg[ \pi^{\top}_{\theta}(o) \Gamma_{\epsilon_k}(o)  - D_k \bigg].
\end{align}
Note that the updates for the reward critic and the entropy multiplier $\alpha$ remain the same as in \alg{PRIMAL-Avg}. Clearly, by replacing the standard cost critic with a distributional one and optimizing against $\mathrm{CVaR}$, \alg{PRIMAL-CVaR} provides a principled framework for building a robust routing policy that is explicitly sensitive to tail-end risks. This is critical for ensuring reliable performance in highly dynamic LEO networks. To summarize our proposed \alg{PRIMAL} algorithm, we present its pseudo code in Algorithm \ref{alg:primal} which integrates the two variants together for clarity. Note that we use a shared centralized replay buffer here by following the Centralized Training and Decentralized Execution (CTDE) paradigm during offline training. However, it can be easily extended to use private replay buffers via online federated learning, as shown in \cite{lozano2025continual}.

\begin{figure}[t!]
    \centering
    \includegraphics[scale=0.32]{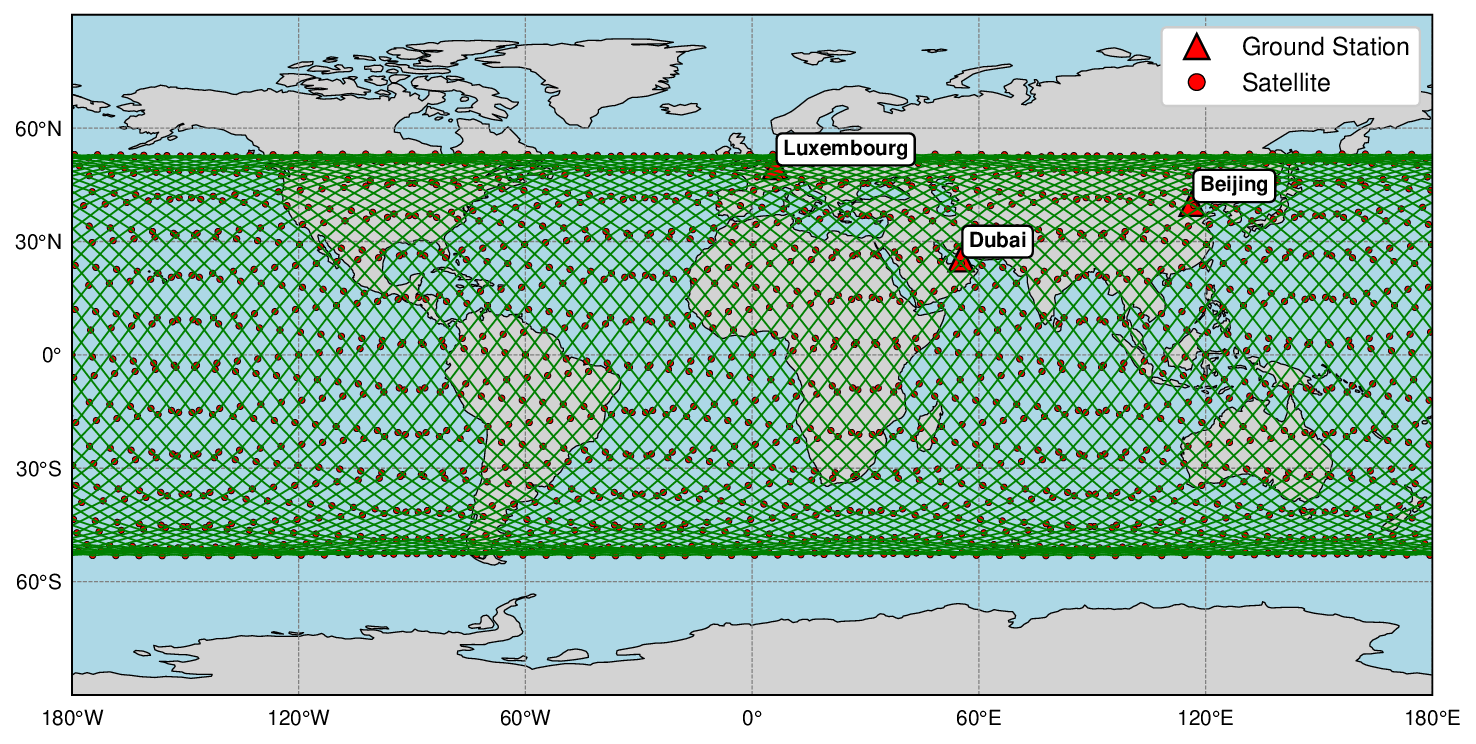}
    \caption{Network topology used in simulations.}
    \label{fig:topology}
\end{figure}

\section{Experiment}
\subsection{Environmental Settings}
To empirically validate our \alg{PRIMAL} framework, we developed an asynchronous event-driven high-fidelity simulator using Python and PyTorch \footnote{Our code is available at \url{https://github.com/skypitcher/risk_aware_marl}}. Specifically, as shown in Fig. \ref{fig:topology}, we run simulations in an ultra-dense Walker-Delta LEO satellite with $22$ satellites per orbit, and $72$ evenly distributed orbits at the same altitude of $600$ km in this Starlink-like constellation ($1584$ satellites). The inclination is $53^\circ$ and the minimum elevation angle is $15^\circ$. We generate packets from three ground stations representing three cities on Earth: Luxembourg, Dubai, and Beijing. All cities have equal probability to be chosen as the source or destination node of a generated packet. We update the satellite positions every $100$ ms. We set stable link data rates at $1000$ Mbps for GSLs and $50$ Mbps for ISLs. The node and link buffer sizes are both $16$ Mbits. We set the traffic packet length following the same setting as in \cite{lozano2025continual}, with $80\%$ being normal packets ($64.8$ Kbits) and the remaining $20\%$ being small packets ($16.2$ Kbits). In addition, we set the maximum TTL as $H=64$. We run the simulation by a $30$-second training or evaluation epoch with a packet traffic rate of $10,000$ packets/s, totaling $300,000$ packets per run according to the Poisson process. We run training iteration once every $1$ ms, and report training performance metrics every $2$ seconds ($2$K iterations).

\begin{figure*}[t!]
    \centering
    \resizebox{0.95\textwidth}{!}{
    \begin{minipage}{0.32\textwidth}
        \centering
        \includegraphics[width=\linewidth]{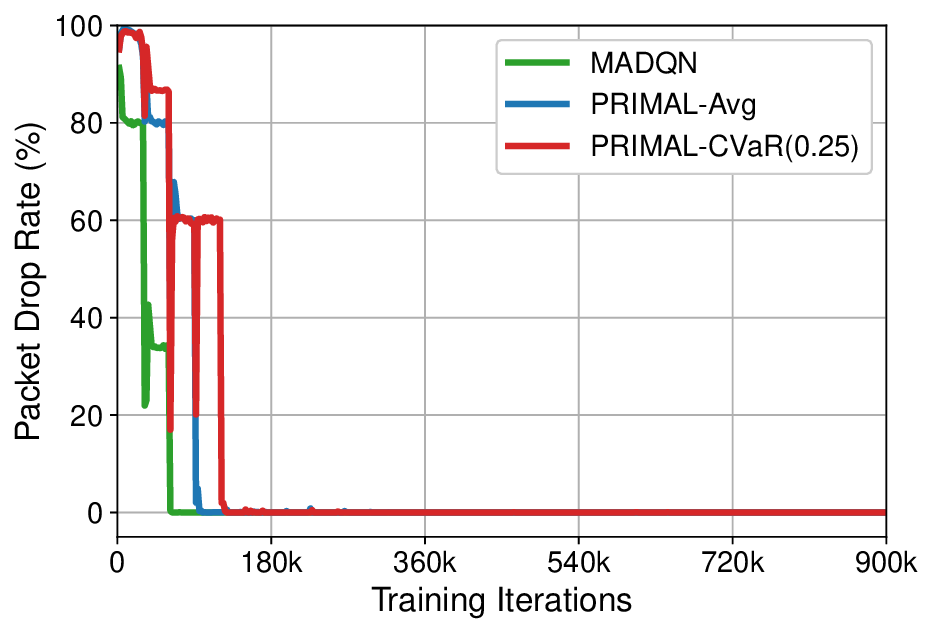}
        \caption{Average packet drop rate versus training epochs for RL algorithms}
        \label{fig:train_drop_rate}
    \end{minipage}
    \hfill
    \begin{minipage}{0.32\textwidth}
        \centering
        \includegraphics[width=\linewidth]{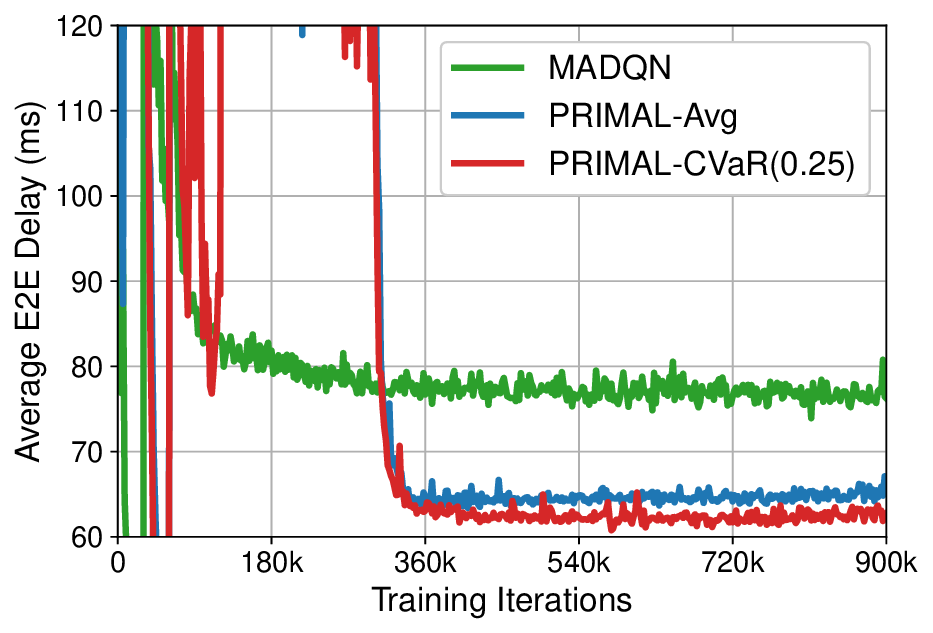}
        \caption{Average E2E delay versus training epochs for RL algorithms}
        \label{fig:train_e2e_delay}
    \end{minipage}
    \hfill
    \begin{minipage}{0.32\textwidth}
        \centering
        \includegraphics[width=\linewidth]{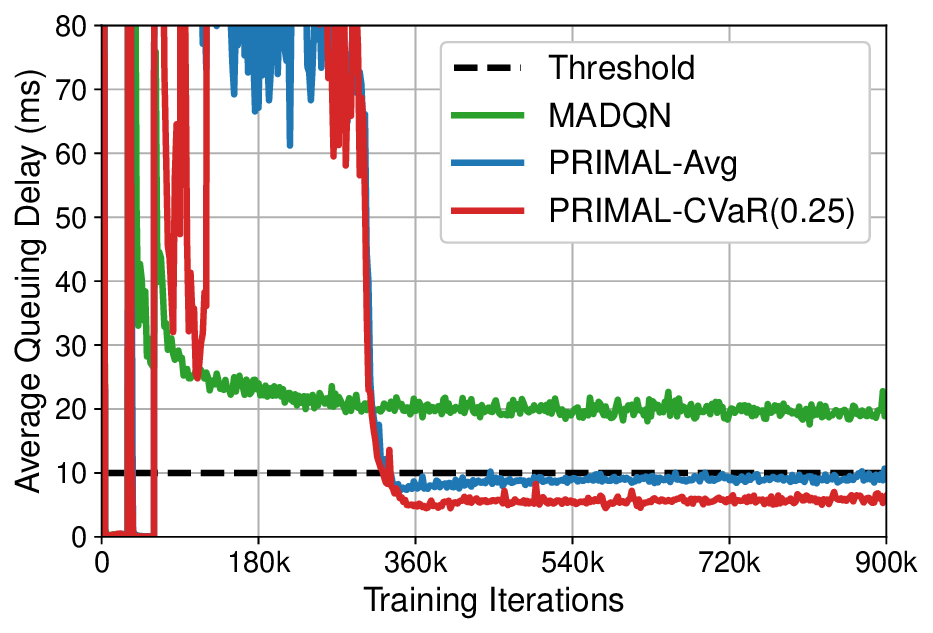}
        \caption{Average queuing delay versus training epochs for RL algorithms}
        \label{fig:train_queuing_delay_avg}
    \end{minipage}
    }
\end{figure*}

\begin{figure*}[t!]
    \centering
    \resizebox{0.95\textwidth}{!}{
    \begin{minipage}{0.32\textwidth}
        \centering
        \includegraphics[width=\linewidth]{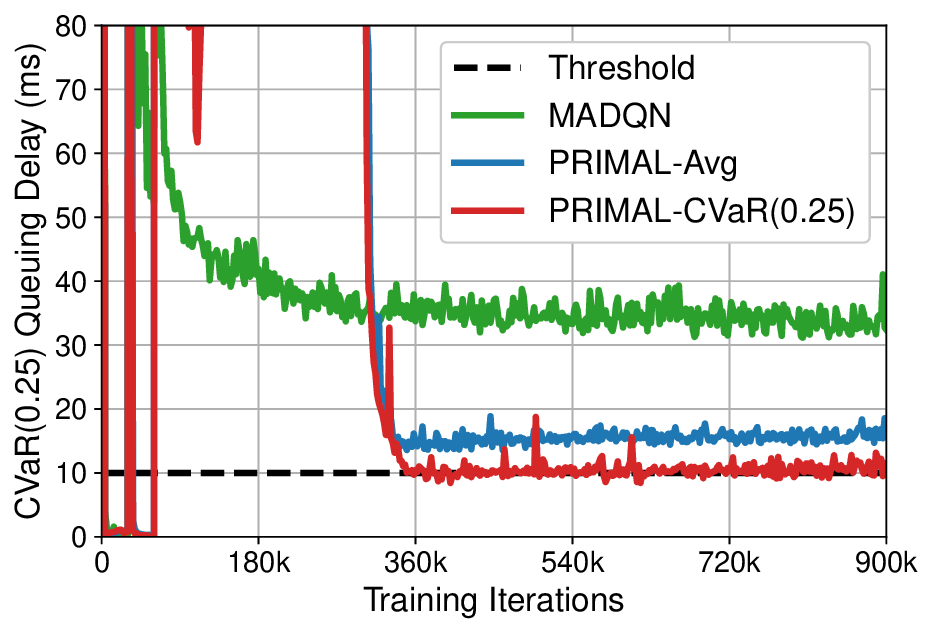}
        \caption{$\mathrm{CVaR}_{0.25}$ Queuing delays versus training epochs for RL algorithms}
        \label{fig:train_queuing_delay_cvar}
    \end{minipage}
    \hfill
    \begin{minipage}{0.32\textwidth}
        \centering
        \includegraphics[width=\linewidth]{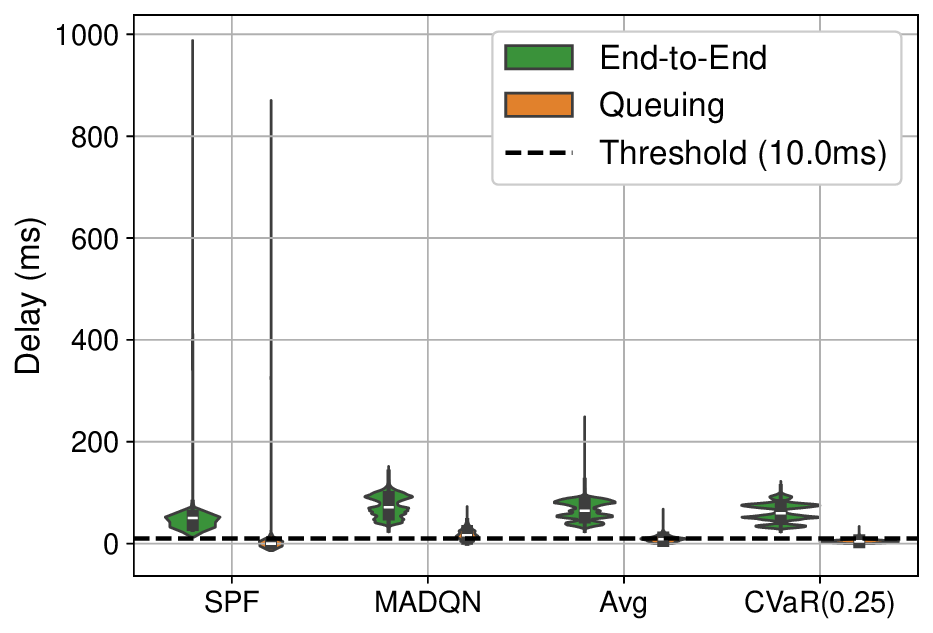}
        \caption{Evaluated delay distribution comparison for the competing algorithms}
        \label{fig:test_delay_violin}
    \end{minipage}
    \hfill
    \begin{minipage}{0.32\textwidth}
        \centering
        \includegraphics[width=\linewidth]{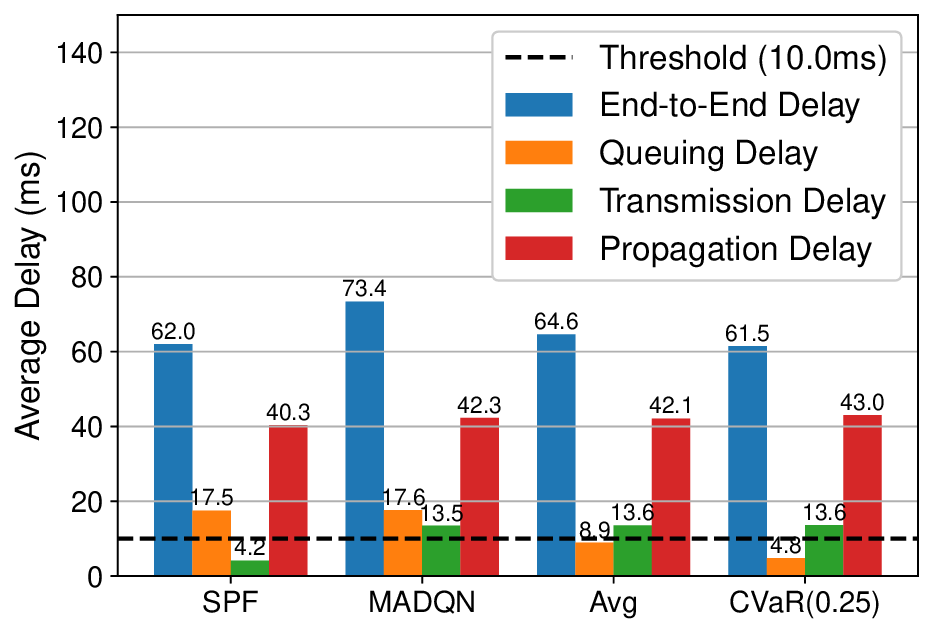}
        \caption{Evaluated delay components comparison for the competing algorithms}
        \label{fig:test_delay_bar}
    \end{minipage}
    }
\end{figure*}

For the neural network implementation, the actor ($\pi_\theta$) and critics ($Q^r_\phi$, $Q^c_{\psi_k}$) use a shared backbone, a two-layer MLP with 512 hidden units, to process observations. Each component has a separate MLP output head. For the \alg{PRIMAL-CVaR} variant, the cost critic $Q^c_{\psi_k}$ is an IQN suggested in \cite{icml/DabneyOSM18} with two layers and quantile parameters $N=N'=N^k=64$. All algorithms use a mini-batch size of $1024$, a replay buffer size of $300000$, and discount factors $\gamma_r=0.99$ and $\gamma_c=0.97$.

For the cost and reward functions, we define the cost function as the normalized queuing delay $c_h = D^{Q}_{h}/D_{norm}$, where $D^{Q}_{h}$ is the queuing delay experienced when packet $p$ is forwarded over a link at hop $h$, and $D_{norm}=100$ ms is a predefined constant for normalization. This cost directly quantifies the level of local congestion. The reward function is designed to minimize delay and maximize delivery rates by combining a dense progressive reward and a terminal reward $B_p$. The per-hop reward is defined as:
\begin{align}
    r_h = \frac{\tau}{D_{norm}} - c_h + \Delta d + B_p
\end{align}
where $\tau$ is the total action delay, and $\Delta d$ provides a dense reward for geographic progress towards the destination, which measured by the difference of Great Circle Distance (GCD). A large terminal reward $B_p$ is added at the final hop to prioritize successful packet delivery:
\begin{align}
    B_p =
  \begin{cases}
    1 + L_p, & \text{if $p$ is delivered,} \\
    -\frac{5\tau^{ttl}_p}{D_{norm}} - \sum^h_{j=0} \Delta d_{GCD}, & \text{if $p$ is dropped,} \\
    {0}, & \text{else,}
  \end{cases}
\end{align}
which encourages maximizing the packet delivery rate. For comparison, we use a hand-crafted reward
\begin{align}\label{eq:hand_reward}
    \bar{r}(o,a,o')=r(o,a,o')+\sum^K_{k=1} c_k(o,a,o')
\end{align}
as the reward function used by heuristic reward shaping approaches, which is equivalent to $\lambda_k=1$ as all these components are well normalized. In addition, we set the minimum entropy as $\bar{\mathcal{H}} \approx 0.067$, which is a heuristic value considering the best action confidence to be $0.99$ while the rest actions have the same probability of $\frac{1-0.99}{|\mathcal{A}|-1}$.

In all simulations, we set the threshold for the queuing delay cost as $D^Q_{max}=10$ ms in total for each packet, and we compare the following algorithms:
\begin{itemize}
    \item \alg{SPF}: The Dijkstra's Shortest Path First (SPF) algorithm, assuming that routing table as precomputed based on the predictable orbital movements.
    \item \alg{MADQN}: The multi-agent asynchronous DQN proposed in \cite{lozano2025continual}, using \eqref{eq:hand_reward} as the effective reward function that relies on human prior knowledge.
    \item \alg{PRIMAL-Avg} and \alg{PRIMAL-CVaR}($\epsilon$): The two variants introduced in this paper, with a risk level $\epsilon \in [0,1]$. Note that \cite{jsac/LyuHFLAM24} uses a similar expectation based cost critic as in \alg{PRIMAL-Avg}. However, \cite{jsac/LyuHFLAM24} is not asynchronous and requires impractical synchronized joint-actions. While it can not work in our asynchronous simulator, we can treat \alg{PRIMAL-Avg} as an asynchronous alternative for it.
\end{itemize}

\subsection{Simulation Results}
We now presents the empirical evaluation of our proposed PRIMAL algorithm against baseline methods. Figs.\ref{fig:train_drop_rate}-\ref{fig:train_queuing_delay_cvar} illustrate the training performance of the reinforcement learning agents. Specifically, Fig. \ref{fig:train_drop_rate} shows that all learning-based algorithms, \alg{MADQN}, \alg{PRIMAL-Avg}, and \alg{PRIMAL-CVaR}, quickly learn to minimize the packet drop rate, achieving an almost-zero drop rate after approximately $150$K training iterations. This indicates that all agents successfully learn the primary objective of delivering packets to their destinations. However, the algorithms show significant differences in their ability to manage network delay and constraints. As shown in Fig. \ref{fig:train_e2e_delay}, while all agents reduce the E2E packet delay over time, the \alg{PRIMAL} variants achieve significantly better performance. \alg{PRIMAL-CVaR} converges to the lowest average E2E delay of approximately $62$ ms, followed by \alg{PRIMAL-Avg} at around $66$ ms, whereas \alg{MADQN} stabilizes at a higher delay of about $77$ ms.

\begin{table*}[t!]
\centering
\caption{Routing performance comparison for the competing algorithms.}
\label{table:performance_comparison}
\begin{tabular}{|l|r|r|c|cr|rc|}
    \hline
    & \multirow{2}{*}{Throughput} & \multirow{2}{*}{Drop Rate} & \multirow{2}{*}{E2E Delay} & \multicolumn{2}{c|}{Queuing Delay} & \multicolumn{2}{c|}{Load Balancing Constraint} \\
    & & & & mean $\pm$ std & $\mathrm{CVaR}_{0.25}$ & Violation Rate\tablefootnote{Dropped packets are also included} & Magnitude of violation\tablefootnote{Only counts for delivered packets} \\
    \hline
    \alg{SPF} & 27.0 Mbps & 84.8\% & 62.0 $\pm$ 85.0 ms & \viola{17.5} $\pm$ 80.2 ms & \viola{70.1 ms} & 85.7\% & 261.12 $\pm$ 176.60 ms \\
    
    \alg{MADQN} & 542.7 Mbps &  \underline{\textbf{0.00\%}} & 73.4 $\pm$ 20.4 ms & \viola{17.6} $\pm$ 10.1 ms & \viola{31.1 ms} & 75.5\% & 11.60 $\pm$ 8.10 ms \\
    
    \alg{PRIMAL-Avg} & 542.9 Mbps & \underline{\textbf{0.00\%}} & 64.6 $\pm$ 17.7 ms & \satisfy{8.9} $\pm$ 5.3 ms & \viola{16.0 ms} & 38.6\% & 4.19 $\pm$ 3.35 ms \\
    
    \alg{PRIMAL-CVaR} & \underline{\textbf{543.0} Mbps} & \underline{\textbf{0.00\%}} & \underline{\textbf{61.5 $\pm$ 18.2} ms} & \underline{\satisfy{4.8} $\pm$ 3.0 ms} & \underline{\satisfy{8.9 ms}} & \underline{\textbf{5.8\%}} & \underline{\textbf{2.47 $\pm$ 2.38} ms} \\
    \hline
\end{tabular}
\end{table*}

\begin{figure*}[t!]
    \centering
    \resizebox{0.9\textwidth}{!}{
    \begin{tikzpicture}[font=\sffamily]
        \draw[->, thick] (0,0) -- (30,0) node[anchor=north east] {Queuing Delay (ms)};
        \draw[->, thick] (0,0) -- (0,5) node[anchor=south west] {Probability Density};
        \foreach \x in {0,5,10,15,20,25}
            \draw (\x, -0.1) -- (\x, 0.1) node[below=2pt] {\x};        

        \draw[blue, very thick, fill=blue!20, opacity=0.8] plot[smooth, tension=0.0] coordinates {
            (0.0,0.0) (0.0,1.073327) (0.2,1.267354) (0.4,1.296451) (0.6,1.310327) (0.8,1.379271) (1.0,1.472388) (1.2,1.543582) (1.4,1.575560) (1.6,1.577781) (1.8,1.567759) (2.0,1.555688) (2.2,1.542204) (2.4,1.525803) (2.6,1.509778) (2.8,1.502000) (3.0,1.508765) (3.2,1.530824) (3.4,1.568042) (3.6,1.622809) (3.8,1.693281) (4.0,1.769120) (4.2,1.837635) (4.4,1.891680) (4.6,1.932260) (4.8,1.964502) (5.0,1.991381) (5.2,2.012391) (5.4,2.027231) (5.6,2.038297) (5.8,2.049075) (6.0,2.062116) (6.2,2.079145) (6.4,2.101209) (6.6,2.128329) (6.8,2.160478) (7.0,2.198815) (7.2,2.244734) (7.4,2.298420) (7.6,2.359234) (7.8,2.426472) (8.0,2.499233) (8.2,2.576444) (8.4,2.655649) (8.6,2.729891) (8.8,2.789392) (9.0,2.829914) (9.2,2.855822) (9.4,2.870743) (9.6,2.869736) (9.8,2.845866) (10.0,2.797797) (10.2,2.727075) (10.4,2.636958) (10.6,2.534350) (10.8,2.426898) (11.0,2.318764) (11.2,2.208633) (11.4,2.093104) (11.6,1.974443) (11.8,1.860836) (12.0,1.758401) (12.2,1.667767) (12.4,1.586588) (12.6,1.512178) (12.8,1.443438) (13.0,1.380953) (13.2,1.324492) (13.4,1.271998) (13.6,1.222233) (13.8,1.176598) (14.0,1.137461) (14.2,1.104865) (14.4,1.075211) (14.6,1.044744) (14.8,1.013107) (15.0,0.981621) (15.2,0.950975) (15.4,0.921700) (15.6,0.894222) (15.8,0.868290) (16.0,0.843241) (16.2,0.818518) (16.4,0.793835) (16.6,0.769170) (16.8,0.744833) (17.0,0.721258) (17.2,0.698263) (17.4,0.674785) (17.6,0.650274) (17.8,0.625624) (18.0,0.601430) (18.2,0.576513) (18.4,0.550112) (18.6,0.524005) (18.8,0.500383) (19.0,0.479035) (19.2,0.457538) (19.4,0.433812) (19.6,0.408571) (19.8,0.384483) (20.0,0.362846) (20.2,0.342424) (20.4,0.321950) (20.6,0.302147) (20.8,0.284277) (21.0,0.268132) (21.2,0.252376) (21.4,0.236179) (21.6,0.220091) (21.8,0.205114) (22.0,0.191358) (22.2,0.178541) (22.4,0.167097) (22.6,0.157385) (22.8,0.148566) (23.0,0.139413) (23.2,0.129625) (23.4,0.119776) (23.6,0.110353) (23.8,0.101580) (24.0,0.093678) (24.2,0.086530) (24.4,0.079863) (24.6,0.073766) (24.8,0.068303) (25.0,0.063157) (25.2,0.058021) (25.4,0.052905) (25.6,0.048101) (25.8,0.043946) (26.0,0.040526) (26.2,0.037674) (26.4,0.035157) (26.6,0.032740) (26.8,0.030212) (27.0,0.027606) (27.2,0.025185) (27.4,0.022996) (27.6,0.020868) (27.8,0.018812) (28.0,0.016893)
        } -- (30.0,0) -- (0,0) -- cycle;
        
        \draw[green!50!black, very thick, fill=green!30, opacity=0.7] plot[smooth, tension=0.0] coordinates {
            (0.0,0.0) (0.0,1.953535) (0.2,1.950051) (0.4,1.687534) (0.6,1.861476) (0.8,2.197784) (1.0,2.564283) (1.2,2.857422) (1.4,3.053521) (1.6,3.280170) (1.8,3.522683) (2.0,3.751229) (2.2,3.989421) (2.4,4.207401) (2.6,4.342410) (2.8,4.441929) (3.0,4.572521) (3.2,4.721738) (3.4,4.860035) (3.6,4.987091) (3.8,5.000000) (4.0,4.890687) (4.2,4.789889) (4.4,4.715935) (4.6,4.643909) (4.8,4.566784) (5.0,4.441099) (5.2,4.255858) (5.4,4.075584) (5.6,3.911156) (5.8,3.746235) (6.0,3.586141) (6.2,3.426247) (6.4,3.235441) (6.6,3.023519) (6.8,2.824258) (7.0,2.645219) (7.2,2.475028) (7.4,2.310675) (7.6,2.151144) (7.8,1.982298) (8.0,1.817288) (8.2,1.676752) (8.4,1.553668) (8.6,1.440323) (8.8,1.329779) (9.0,1.214502) (9.2,1.112918) (9.4,1.024378) (9.6,0.936300) (9.8,0.855410) (10.0,0.791077) (10.2,0.733829) (10.4,0.672904) (10.6,0.614865) (10.8,0.561053) (11.0,0.516750) (11.2,0.475425) (11.4,0.436208) (11.6,0.401260) (11.8,0.371273) (12.0,0.343105) (12.2,0.313744) (12.4,0.285117) (12.6,0.261819) (12.8,0.244067) (13.0,0.225059) (13.2,0.204630) (13.4,0.188347) (13.6,0.174242) (13.8,0.158648) (14.0,0.147143) (14.2,0.136670) (14.4,0.126961) (14.6,0.118471) (14.8,0.111321) (15.0,0.104958) (15.2,0.098091) (15.4,0.090586) (15.6,0.083761) (15.8,0.077076) (16.0,0.071827) (16.2,0.068515) (16.4,0.064776) (16.6,0.060982) (16.8,0.056941) (17.0,0.052712) (17.2,0.049461) (17.4,0.047488) (17.6,0.044954) (17.8,0.041646) (18.0,0.037678) (18.2,0.034527) (18.4,0.033351) (18.6,0.031432) (18.8,0.028395) (19.0,0.025985) (19.2,0.023228) (19.4,0.020996) (19.6,0.019264) (19.8,0.016779) (20.0,0.014245) (20.2,0.011886) (20.4,0.010871) (20.6,0.010228) (20.8,0.009183) (21.0,0.008165) (21.2,0.007296) (21.4,0.006041) (21.6,0.004651) (21.8,0.004344) (22.0,0.004574) (22.2,0.003998) (22.4,0.002819) (22.6,0.002313) (22.8,0.002189) (23.0,0.001952) (23.2,0.001815) (23.4,0.001861) (23.6,0.001692) (23.8,0.001310) (24.0,0.001042) (24.2,0.000969) (24.4,0.000948) (24.6,0.000707) (24.8,0.000456) (25.0,0.000452) (25.2,0.000585) (25.4,0.000640) (25.6,0.000639) (25.8,0.000645) (26.0,0.000534) (26.2,0.000370)
        } -- (30.0,0) -- (0,0) -- cycle;
    
        \begin{scope}
            \clip (10, 0) rectangle (28, 5);
            \fill[pattern=north east lines, pattern color=red!80] plot[smooth, tension=0.0] coordinates {
                 (0.0,0.0) (0.0,1.073327) (0.2,1.267354) (0.4,1.296451) (0.6,1.310327) (0.8,1.379271) (1.0,1.472388) (1.2,1.543582) (1.4,1.575560) (1.6,1.577781) (1.8,1.567759) (2.0,1.555688) (2.2,1.542204) (2.4,1.525803) (2.6,1.509778) (2.8,1.502000) (3.0,1.508765) (3.2,1.530824) (3.4,1.568042) (3.6,1.622809) (3.8,1.693281) (4.0,1.769120) (4.2,1.837635) (4.4,1.891680) (4.6,1.932260) (4.8,1.964502) (5.0,1.991381) (5.2,2.012391) (5.4,2.027231) (5.6,2.038297) (5.8,2.049075) (6.0,2.062116) (6.2,2.079145) (6.4,2.101209) (6.6,2.128329) (6.8,2.160478) (7.0,2.198815) (7.2,2.244734) (7.4,2.298420) (7.6,2.359234) (7.8,2.426472) (8.0,2.499233) (8.2,2.576444) (8.4,2.655649) (8.6,2.729891) (8.8,2.789392) (9.0,2.829914) (9.2,2.855822) (9.4,2.870743) (9.6,2.869736) (9.8,2.845866) (10.0,2.797797) (10.2,2.727075) (10.4,2.636958) (10.6,2.534350) (10.8,2.426898) (11.0,2.318764) (11.2,2.208633) (11.4,2.093104) (11.6,1.974443) (11.8,1.860836) (12.0,1.758401) (12.2,1.667767) (12.4,1.586588) (12.6,1.512178) (12.8,1.443438) (13.0,1.380953) (13.2,1.324492) (13.4,1.271998) (13.6,1.222233) (13.8,1.176598) (14.0,1.137461) (14.2,1.104865) (14.4,1.075211) (14.6,1.044744) (14.8,1.013107) (15.0,0.981621) (15.2,0.950975) (15.4,0.921700) (15.6,0.894222) (15.8,0.868290) (16.0,0.843241) (16.2,0.818518) (16.4,0.793835) (16.6,0.769170) (16.8,0.744833) (17.0,0.721258) (17.2,0.698263) (17.4,0.674785) (17.6,0.650274) (17.8,0.625624) (18.0,0.601430) (18.2,0.576513) (18.4,0.550112) (18.6,0.524005) (18.8,0.500383) (19.0,0.479035) (19.2,0.457538) (19.4,0.433812) (19.6,0.408571) (19.8,0.384483) (20.0,0.362846) (20.2,0.342424) (20.4,0.321950) (20.6,0.302147) (20.8,0.284277) (21.0,0.268132) (21.2,0.252376) (21.4,0.236179) (21.6,0.220091) (21.8,0.205114) (22.0,0.191358) (22.2,0.178541) (22.4,0.167097) (22.6,0.157385) (22.8,0.148566) (23.0,0.139413) (23.2,0.129625) (23.4,0.119776) (23.6,0.110353) (23.8,0.101580) (24.0,0.093678) (24.2,0.086530) (24.4,0.079863) (24.6,0.073766) (24.8,0.068303) (25.0,0.063157) (25.2,0.058021) (25.4,0.052905) (25.6,0.048101) (25.8,0.043946) (26.0,0.040526) (26.2,0.037674) (26.4,0.035157) (26.6,0.032740) (26.8,0.030212) (27.0,0.027606) (27.2,0.025185) (27.4,0.022996) (27.6,0.020868) (27.8,0.018812) (28.0,0.016893)
            } -- (30.0,0) -- (0,0) -- cycle;
        \end{scope}
        
        \begin{scope}
            \clip (10, 0) rectangle (28, 5);
             \fill[pattern=crosshatch, pattern color=green!80!black] plot[smooth, tension=0.0] coordinates {
                (0.0,0.0) (0.0,1.953535) (0.2,1.950051) (0.4,1.687534) (0.6,1.861476) (0.8,2.197784) (1.0,2.564283) (1.2,2.857422) (1.4,3.053521) (1.6,3.280170) (1.8,3.522683) (2.0,3.751229) (2.2,3.989421) (2.4,4.207401) (2.6,4.342410) (2.8,4.441929) (3.0,4.572521) (3.2,4.721738) (3.4,4.860035) (3.6,4.987091) (3.8,5.000000) (4.0,4.890687) (4.2,4.789889) (4.4,4.715935) (4.6,4.643909) (4.8,4.566784) (5.0,4.441099) (5.2,4.255858) (5.4,4.075584) (5.6,3.911156) (5.8,3.746235) (6.0,3.586141) (6.2,3.426247) (6.4,3.235441) (6.6,3.023519) (6.8,2.824258) (7.0,2.645219) (7.2,2.475028) (7.4,2.310675) (7.6,2.151144) (7.8,1.982298) (8.0,1.817288) (8.2,1.676752) (8.4,1.553668) (8.6,1.440323) (8.8,1.329779) (9.0,1.214502) (9.2,1.112918) (9.4,1.024378) (9.6,0.936300) (9.8,0.855410) (10.0,0.791077) (10.2,0.733829) (10.4,0.672904) (10.6,0.614865) (10.8,0.561053) (11.0,0.516750) (11.2,0.475425) (11.4,0.436208) (11.6,0.401260) (11.8,0.371273) (12.0,0.343105) (12.2,0.313744) (12.4,0.285117) (12.6,0.261819) (12.8,0.244067) (13.0,0.225059) (13.2,0.204630) (13.4,0.188347) (13.6,0.174242) (13.8,0.158648) (14.0,0.147143) (14.2,0.136670) (14.4,0.126961) (14.6,0.118471) (14.8,0.111321) (15.0,0.104958) (15.2,0.098091) (15.4,0.090586) (15.6,0.083761) (15.8,0.077076) (16.0,0.071827) (16.2,0.068515) (16.4,0.064776) (16.6,0.060982) (16.8,0.056941) (17.0,0.052712) (17.2,0.049461) (17.4,0.047488) (17.6,0.044954) (17.8,0.041646) (18.0,0.037678) (18.2,0.034527) (18.4,0.033351) (18.6,0.031432) (18.8,0.028395) (19.0,0.025985) (19.2,0.023228) (19.4,0.020996) (19.6,0.019264) (19.8,0.016779) (20.0,0.014245) (20.2,0.011886) (20.4,0.010871) (20.6,0.010228) (20.8,0.009183) (21.0,0.008165) (21.2,0.007296) (21.4,0.006041) (21.6,0.004651) (21.8,0.004344) (22.0,0.004574) (22.2,0.003998) (22.4,0.002819) (22.6,0.002313) (22.8,0.002189) (23.0,0.001952) (23.2,0.001815) (23.4,0.001861) (23.6,0.001692) (23.8,0.001310) (24.0,0.001042) (24.2,0.000969) (24.4,0.000948) (24.6,0.000707) (24.8,0.000456) (25.0,0.000452) (25.2,0.000585) (25.4,0.000640) (25.6,0.000639) (25.8,0.000645) (26.0,0.000534) (26.2,0.000370)
            } -- (30.0,0) -- (0,0) -- cycle;
        \end{scope}
        
        \draw[->, blue, thick] (12.7, 3.5) -- (9.0, 2.0);
        \node[blue, align=left] at (14.5, 3.5) {Average $\approx$ 8.9ms \\ (Satisfies Threshold)};
    
        \draw[->, blue, thick] (22, 1.5) -- (16.1, 0.5);
        \node[red, align=left] at (22, 2.0) {$\mathrm{CVaR}_{0.25}$ $\approx$ 16.0ms \\(Violates Threshold)};

        \draw[dashed, blue, very thick] (8.9,0) -- (8.9,2.818597) node[pos=0.9, right, text=red] {};
        \draw[dashed, blue, very thick] (16.0,0) -- (16.0,0.843051) node[pos=0.9, right, text=red] {};
        
        \draw[->, black!50!black, thick] (3.7, 0.7) -- (4.7, 0.5);
        \node[black!50!black, align=right] at (2, 0.5) {Average $\approx$ 4.8ms \\ (Satisfies Threshold)};
    
        \draw[->, black!50!black, thick] (6.7, 0.8) -- (8.7, 0.25);
        \node[black!50!black, align=right] at (6.7, 1.2) {$\mathrm{CVaR}_{0.25}$ $\approx$ 8.9ms \\ (Satisfies Threshold)};
        
        \draw[dashed, black, very thick] (4.8,0) -- (4.8,4.557992) node[pos=0.9, right, text=red] {};
        \draw[dashed, black, very thick] (8.9,0) -- (8.9,1.274490) node[pos=0.9, right, text=red] {};
        
        \draw[dashed, red, ultra thick] (10,0) -- (10,6) node[pos=0.9, right, text=red] {Threshold (10ms)};
        
        \node[draw, anchor=north east, align=left] at (29.1, 5.5) {
            \begin{tabular}{ll}
            \tikz\draw[blue, fill=blue!20, very thick] (0,0) rectangle (0.4,0.4); & \alg{PRIMAL-Avg} (Risk-Myopic) \\
            \tikz\draw[green!50!black, fill=green!30, very thick] (0,0) rectangle (0.4,0.4); & \alg{PRIMAL-CVaR}(0.25) (Risk-Averse) \\
            \end{tabular}
        };
        
        \node[draw, fill=yellow!10, align=left, text width=12cm] at (23, 3.2) {
            \alg{PRIMAL-Avg} exhibits a fat distribution with a seriously heavier tail than \alg{PRIMAL-CVaR}, which clearly illustrates the risk-myopic nature of expectation based cost critics.
        };
    
    \end{tikzpicture}
    }
    \caption{Learned policy's queuing delay distribution comparison for the proposed \textbf{\texttt{PRIMAL}} variants. Data were directly drawn from well-trained models.}
    \label{fig:policy_queuing_delay_pdf_comparison}
\end{figure*}
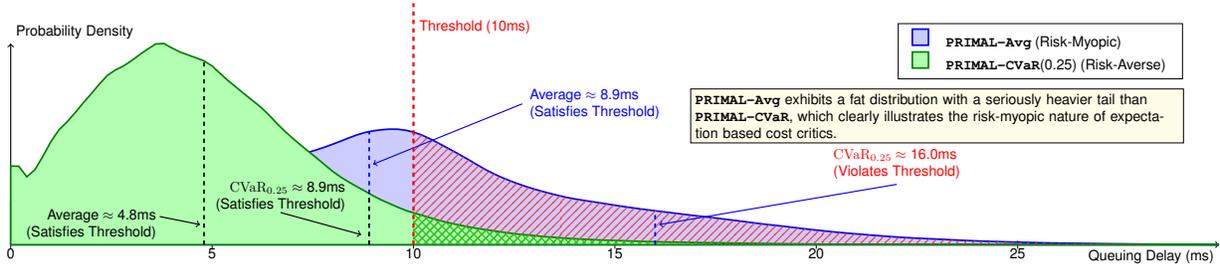

The difference in performance is largely explained by how each algorithm handles the queuing delay constraint, which is a direct indicator for network congestion. Fig. \ref{fig:train_queuing_delay_avg} demonstrates that \alg{MADQN} fails to respect the $10$ ms queuing delay threshold, converging to a value consistently near $18$ ms. In contrast, \alg{PRIMAL-Avg}, which directly optimizes for the expected cost, successfully learns to keep the queuing delay at the $10$ ms threshold. \alg{PRIMAL-CVaR} goes one step further, reducing the average queuing delay to just $5$ ms well below the threshold, by actively mitigating worst-case tail events. The unique capability of \alg{PRIMAL-CVaR} is clearly validated in Fig. \ref{fig:train_queuing_delay_cvar}, which plots the $\mathrm{CVaR}_{0.25}$ of the queuing delay. We can find that only \alg{PRIMAL-CVaR} successfully reduces the tail-end risk, bringing the $\mathrm{CVaR}_{0.25}$ of the queuing delay down to the $10$ ms threshold, whereas both \alg{MADQN} and \alg{PRIMAL-Avg} exhibit a $\mathrm{CVaR}_{0.25}$ far exceeding this limit. This strongly confirms that \alg{PRIMAL-CVaR} is highly risk-aware and is capable to effectively constrain worst-case congestion events.

The post-training evaluation results are summarized in Table~\ref{table:performance_comparison} and detailed in Figs.~\ref{fig:test_delay_violin} and \ref{fig:test_delay_bar}, averaged over five test runs with different random seeds. The static baseline \alg{SPF} performs poorly, suffering an $84.8$\% drop rate and high E2E delay variance. This shows its inability to handle dynamic congestion. In contrast, all learning-based methods achieve high throughput and near-zero drop rates. Among them, \alg{PRIMAL-CVaR} delivers the best overall performance: the highest throughput ($543.0$ Mbps), the lowest E2E delay ($61.5 \pm 18.2$ ms), and minimal queuing delay ($4.8 \pm 3.0$ ms). \alg{PRIMAL-Avg} also satisfies the average queuing delay constraint ($8.9 \pm 5.3$ ms) and outperforms \alg{MADQN}, which violates the constraint with $17.6 \pm 10.1$ ms due to its heuristic risk handling. To assess risk, we report $\mathrm{CVaR}_{0.25}$ of queuing delay: \alg{MADQN} and \alg{PRIMAL-Avg} incur $31.1$ ms and $16.0$ ms, while only \alg{PRIMAL-CVaR} keeps it below the $10$ ms threshold at $8.9$ ms, indicating effective tail-risk mitigation. As a result, it also achieves the lowest violation magnitude, which outperforms all other methods.

\footnotetext[2]{Packet dropping is considered as load balance violation as well.}
\footnotetext[3]{We only counts for packets that are successfully delivered.}

Moreover, Figs. \ref{fig:test_delay_violin} and \ref{fig:policy_queuing_delay_pdf_comparison} provide a distributional view of the E2E and queuing delays. The results for SPF and MADQN show very high variance, indicating unpredictable performance. In contrast, the PRIMAL variants, particularly PRIMAL-CVaR, show much tighter delay distributions. In Fig. \ref{fig:policy_queuing_delay_pdf_comparison}, while both the two \alg{PRIMAL} variants keep the average queuing delay below the $10$ms threshold, the distributions reveal a key difference in risk management. \alg{PRIMAL-Avg} exhibits a wider distribution with a heavier tail, and its CVaR at a $25$\% risk level ($\mathrm{CVaR}_{0.25}$) significantly violates the threshold. This demonstrates \alg{PRIMAL-CVaR}'s ability to not only optimize for average performance but to also effectively mitigate high-delay tail events, resulting in a more predictable and robust routing policy.

Fig. \ref{fig:test_delay_bar} breaks down the average end-to-end delay into its constituent parts, offering crucial insights into the operational differences between the routing strategies. From the delay compositions of the \alg{PRIMAL} variants, a well-known (and almost common sense) principle is clearly validated: the fastest path is not always the geographically shortest one. For example, \alg{PRIMAL-CVaR} accepts a minor increase in propagation delay over \alg{PRIMAL-Avg} (only $1$ ms higher), indicating it selects physically longer routes. However, what makes this result interesting is not the validation of the principle, but how our algorithm effectively realizes this well-know trade-off. With the minimal detour cost, \alg{PRIMAL-CVaR} achieves a massive ~$46$\% reduction in queuing delay comparing to \alg{PRIMAL-Avg}, leading to more effective and balanced network load. This highlights that risk-aware congestion avoidance is the most critical factor influencing performance, far outweighing the marginal cost of a slightly longer path. By making this intelligent risk-aware trade-off, \alg{PRIMAL-CVaR} effectively bypasses network hotspots to achieve the lowest overall end-to-end delay with almost-zero packet dropping rate.

\section{Conclusion}
In this paper, we proposed \alg{PRIMAL}, an asynchronous risk-aware multi-agent packet routing framework tailored for the dynamic decentralized nature of LEO satellite networks. Our event-driven design enables each satellite to act independently with their own pace, while risk-awareness is achieved through primal-dual learning using distributional RL to capture routing cost distributions and constrain tail risks via $\mathrm{CVaR}$. Empirical results show that \alg{PRIMAL} effectively avoids traffic hotspots by trading off slightly longer paths for improved load balancing and overall performance, demonstrating a principled approach to risk-aware routing in highly dynamic networks.

\clearpage

\bibliographystyle{IEEEtran}
\bibliography{IEEEabrv, CRN}

\end{document}